\documentclass[preprint, APL]{revtex4}
\usepackage{bm}% bold math
\usepackage{amssymb}
\usepackage{amsmath}
\usepackage[sort&compress]{natbib}
\usepackage{epsfig}
\usepackage[dvips]{color}

\usepackage[colorlinks=true, pagebackref=false, bookmarks=true,bookmarksopen=true,bookmarksnumbered=true]{hyperref}

\newcommand{\Al}{\ensuremath{\mathrm{Al}}}

\newcommand{\N}{\ensuremath{\mathrm{N}}}
\newcommand{\Ti}{\ensuremath{\mathrm{Ti}}}

\newcommand{\AlTiN}{$\Al/\Ti\N$ }

\renewcommand{\section}[1]{}
\renewcommand{\subsection}[1]{}

\begin{document}

\title{Identifying capacitive and inductive loss in lumped element superconducting hybrid titanium nitride/ aluminum resonators\footnote{This contribution of NIST, an agency of the U.S. Government, is not subject to copyright.}}%

\author{Michael R. Vissers}
\email{michael.vissers@NIST.gov}
\affiliation{National Institute of Standards and Technology, Boulder, Colorado 80305, USA}

\author{Martin P. Weides}\altaffiliation{Current address: Karlsruhe Institute of Technology, 76131 Karlsruhe, Germany}

\author{Jeffrey S. Kline}
\affiliation{National Institute of Standards and Technology, Boulder, Colorado 80305, USA}

\author{Martin O. Sandberg}
\affiliation{National Institute of Standards and Technology, Boulder, Colorado 80305, USA}

\author{David P. Pappas}
\email{david.pappas@NIST.gov}
\affiliation{National Institute of Standards and Technology, Boulder, Colorado 80305, USA}

\date{\today}% It is always \today, today,
             % but any date may be explicitly specified

\begin{abstract}
We present a method to systematically locate and extract capacitive and inductive losses in superconducting resonators at microwave frequencies by use of mixed-material, lumped element devices. In these devices, ultra-low loss titanium nitride was progressively replaced with aluminum in the inter-digitated capacitor and meandered inductor elements. By measuring the power dependent loss at $50\;\rm{mK}$ as the \AlTiN fraction in each element is increased, we find that at low electric field, i.e. in the single photon limit, the loss is two level system in nature and is correlated with the amount of $\Al$ capacitance rather than the Al inductance. In the high electric field limit, the remaining loss is linearly related to the product of the Al area times its inductance and is likely due to quasiparticles generated by stray radiation.  At elevated temperature, additional loss is correlated with the amount of Al in the inductance, with a power independent TiN-Al interface loss term that exponentially decreases as the temperature is reduced. The TiN-Al interface loss is vanishingly small at the $50\;\rm{mK}$ base temperature. %This work also highlights the difficulty in making truly lumped element devices, as meansered or interdigitated elements still have considerable self-capacitance or self-inductance. 

\end{abstract}

%Use showkeys class option if keyword display desired

\maketitle

Superconducting resonators with low loss are of great interest for photon detection \cite{DayNature03} and quantum computation \cite{Matteo_Science11}. Hybrid devices, using different materials in specific parts of the resonant circuit, allow the alteration of intrinsic materials properties such as the superconducting gap, radiation cross-section, and capacitive and inductive loss within the resonators to optimize the entire circuit. Locating whether the residual loss sources are inductive or capacitive in nature is of considerable relevance. So far, most resonator studies have been done on single films in either coplanar waveguide resonators (CPWs) or lumped element circuits\cite{Gao_APL08b, OConnellAPL08, BarendsAPL10, BarendsAPL10_2, Khalil, Vissers_APL10,Leduc_APL10}. The distributed capacitive and inductive elements in CPWs complicate the analysis, including the interpretation and location of the measured resonator loss. More specifically, the loss contributions of the capacitive $\delta_C$ and inductive $\delta_L$ parts are inherently linked and cannot be investigated separately in distributed element devices like CPWs. Further, even in lumped element devices typically both the capacitor and the inductor are usually formed by the same material, with limited research done on mixed-material devices with interfaces.   In this work, we address these issues by studying lumped element resonators with components of mixed materials. However, we note that the lumped element approach is an approximation, depending on the exact current and voltage distributions in the resonant circuit. Therefore we distinguish between and investigate the actual capacitive and inductive losses in the specific elements as each element has contributions from both. \par

In a resonator the energy stored is $E_{\rm{total}}$, where on average the capacitive energy equals the inductive energy. The measured resonator loss, $\delta_\textrm{meas.}$, is the measurement of the relative energy loss $\Delta E/E_{\rm{total}}$ per cycle time and defined as 
\begin{equation}
\delta_{\textrm{meas.}}=\frac{\Delta E}{2\pi E_{\rm{total}}}=\frac{\Delta E_C+\Delta E_L}{2\pi E_{\rm{total}}}=\delta_C+\delta_L\;.
\label{eqn:MeasuredLoss}
\end{equation}
The conventional resonator characterization is done by measuring the resonator loss and frequency under power and temperature sweeps respectively. The capacitive loss, $\delta_C$, is believed to be due to the presence of two-level systems (TLSs) \cite{Hunklinger_1976}, and is inferred from both the power dependent loss and resonance frequency temperature dependence, as reviewed in Ref. \cite{Pappas_ASC_11}.  The decreased loss at higher powers is naturally explained by TLS loss, which scales with the electric field, $E$, as 
\begin{equation}
\delta_{\textrm{meas.}} \propto {1 \over \sqrt{1+\left(\frac{E}{E_s}\right)^2}}+ \delta_{\textrm{P.I.}}\;,
\label{eqn:LossVE}
\end{equation} where $E_s$ is a saturation electric field for TLS loss \cite{Hunklinger_1976}. In addition to TLS loss, there is also a power-independent background loss, $\delta_{\textrm{P.I.}}$, typically attributed to the inductive loss, $\delta_L$, caused, for example, by quasiparticles or radiative loss. The power independent loss typically limits the total resonator loss, causing the measured loss' electric field dependence to deviate from $1/E$ at high electric fields \cite{Pappas_ASC_11}.

In this paper, we present a complementary characterization method to determine the capacitive and inductive losses in resonators by use of lumped elements made up of different materials.  Devices fabricated from different materials allow us to vary the $T_c$ and the different microwave loss factors of the constituent components highlighting their contributions to the loss of the different circuit elements. We tested this approach using titanium nitride and aluminum on Si.  In addition to having a $T_c=4.5-5\;\rm{K}$, optimally grown TiN on Si is one of the lowest loss resonator materials with $3$:$2\;\rm{\mu m}$ CPWs having a low electric field, low temperature, loss factor $\sim 1\times10^{-6}$, and high field loss $<1\times10^{-7}$ \cite{Vissers_APL10}.  Evaporated Al, used in qubits e.g. Refs. \onlinecite{MartinisPRL05,Paik_3d_11}, has a $T_c=1.2\;\rm{K}$.  However when grown on Si, the oxide surface and intermixing at the substrate interface give rise to relatively high loss $\delta_\textrm{int} > 10^{-5}$ \cite{WangAPL09}, providing a contrast to the lower loss TiN.  The lower superconducting gap in Al facilitates quasiparticle creation from elevated temperatures or infrared radiation \cite{Corcoles_APL_2011,Barends_Stray_Light}.  In addition to determining the capacitive and inductive loss contributions, we also probe the $\Al\textrm{-}\Ti\N$ material interface, which has not been investigated in previous single layer studies. This interface pears to have a small, but non-zero contribution to the loss at elevated temperature, but negligible at the base temperature.  This work highlights the feasiblity of engineered lumped element resonators made from two or more materials whose specific capacitive or inductive properties are optimally tuned to take advantage of the constituent elements.

For our devices, $100\;\rm{nm}$ thick $\Ti\N$ was first grown at $500 ^{\circ}\rm{C}$ on an intrinsic silicon wafer \cite{Vissers_APL10} and patterned using optical lithography and a Cl based reactive ion etch. While fluorine etched and deeply trenched $\Ti\N$ resonators have one of the lowest loss factors \cite{Vissers_Trench}, the chlorine etch chosen in this work gave a better defined silicon interface without trenches complicating subsequent the liftoff.  However, the Cl based etch does cause slightly increased internal resonator loss of $\delta_\textrm{int.} \approx 5\times 10^{-6}$ \cite{Sandberg_TiN}.  The exposed silicon surface was then radio frequency (RF) plasma cleaned before the aluminum evaporation.  The RF treatment might increase the loss in the exposed silicon and $\Ti\N$ regions. Finally we deposited a $100 \;\rm{nm}$ thick aluminum layer with e-beam evaporation which was patterned with a lift-off process. The $\Al$ film overlapped the $\Ti\N$ film forming $2\;\rm{\mu m}\times2\;\rm{\mu m}$ contacts which were also RF cleaned to give a clean interface.  The films were patterned into frequency multiplexed lumped element resonators with varying degrees of $\Ti\N$ and $\Al$ participation. The internal wiring and microwave measurement lines were made from TiN for all devices. The resonators were capacitively coupled to a microwave coplanar waveguide feedline, permitting the loss extraction from transmission measurments of the $S_{21}$ parameters, and the internal loss of a specific resonator, $\delta_\textrm{int.}$, is given by 
\begin{equation}
\delta_\textrm{int.}=\delta_\textrm{meas.}-\delta_\textrm{coup.}\;,
\label{eqn:InternalLoss}
\end{equation}
where $\delta_\textrm{coup.}$ is the loss due to the capacitive coupling to the feedline.

\begin{figure}[tb]
\begin{center}
\includegraphics[width=8.5cm]{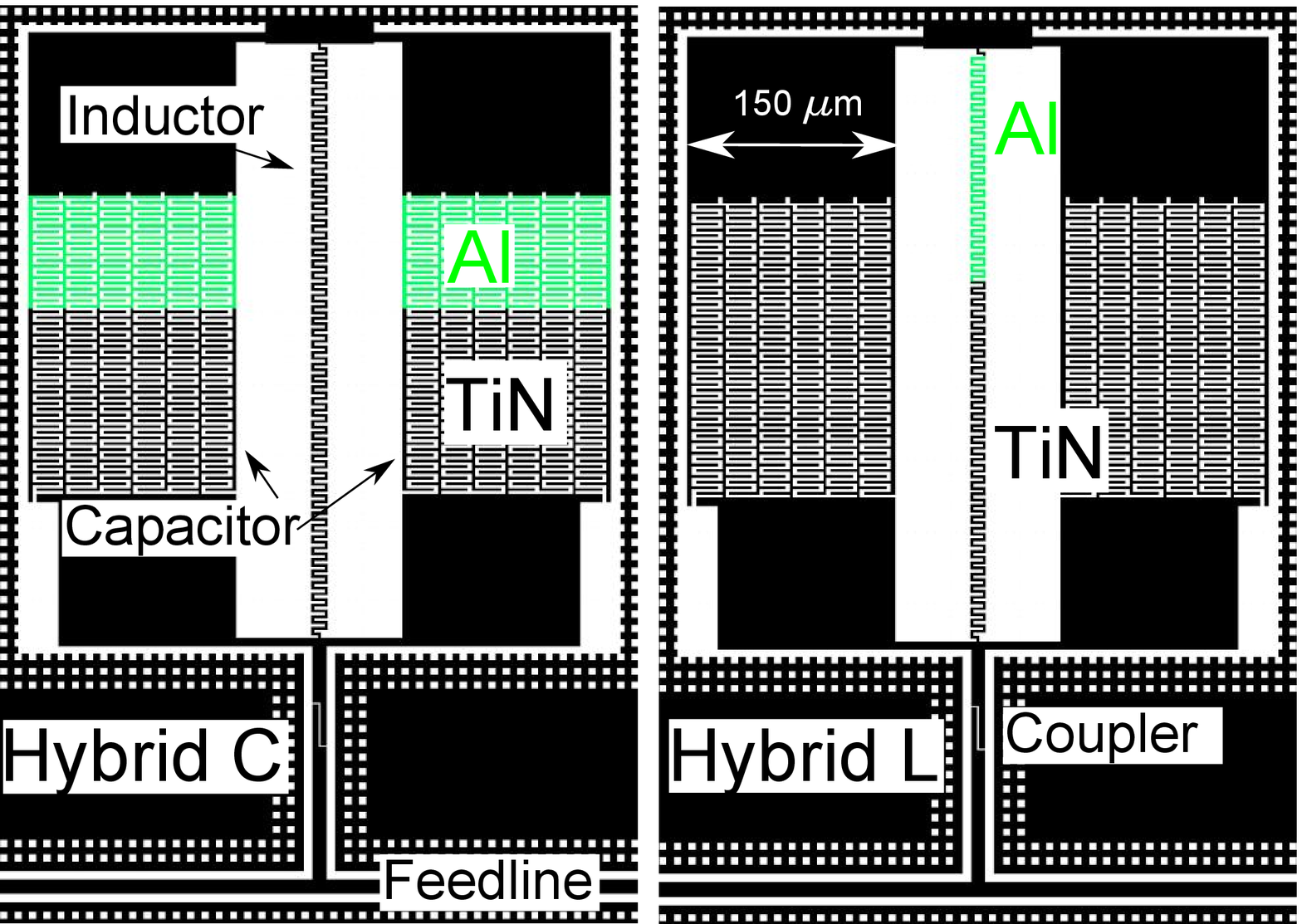}
\caption{(color online)  Drawn resonator designs \emph{hybrid capacitor} (left) and \emph{hybrid inductor} (right) with $3:2\;\Ti\N:\Al$ fraction. The $\Ti\N$ (black) inter-digitated capacitors and meandered inductors are progressively replaced with $\Al$ (green). The frequency multiplexed resonators are capacitively coupled to the feedline.}\label{Fig_1}
\end{center}
\end{figure}

The resonators are formed by an inductor (meandered $2\;\rm{\mu m}$ wide wire with $2\;\rm{\mu m}$ gap) being symmetrically shunted by two inter-digitated capacitors (IDCs) ($2\;\rm{\mu m}$ wide fingers and gaps), see Fig. \ref{Fig_1}. The footprint per resonator is $400\;\rm{\mu m}\times450 \;\rm{\mu m}$. The superconductors' width and gap were chosen to be similar to the features in conventional coplanar waveguide resonators. One resonator electrode is galvanically connected to the ground plane formed by $\Ti\N$. Around the resonator are flux holes in the ground plane to inhibit trapped vortices. This resonator design yields intrinsic loss factors similar to those measured on otherwise identical coplanar waveguide resonators. On the first design, the hybrid capacitor device, the $\Ti\N$ capacitor was progressively replaced with $\Al$ while the inductor was formed by $\Ti\N$. On the second design, the hybrid inductor device, the inductor had varied \AlTiN fractions, while the capacitor was always formed by $\Ti\N$. Comparing simulations with our measurements of the resonant frequency for the different TiN fractions, we determined a kinetic inductance of $0.4\;\rm{pH/square}$ for the TiN film. 

Illustrations of the resonator designs are shown in Figure \ref{Fig_1}. Each design was integrated as frequency multiplexed resonators coupled to a common feedline. In total, for each design, we measured one all-$\Ti\N$, one fully $\Al$ hybrid and four mixed resonant elements with various \AlTiN fractions. The IDC capacitance was designed to be $600\; \rm{fF}$, and the geometric meandered inductance was nominally $0.74 \;\rm{nH}$ for the hybrid inductor devices. Frequency multiplexing of the resonators was a natural consequence of the kinetic inductance of the different $\Ti\N$ fraction. For the hybrid capacitor devices, the geometric inductances were varied ($0.64\textrm{-}0.86\; \rm{nH}$). The self-capacitance of the inductor and inductance of the capacitor plays an important role in the loss of the entire structure.  Using calculations from an electromagnetic field solver we estimate the self-capacitance of the meandered inductor as $18 \%$ of the total capacitance and the self-inductance of the IDC as about $3 \%$ of the total inductance. These calculations agree well with the measured resonance frequencies, $f_r$, of $5.5$ and $6.9\;\rm{GHz}$, though we obtain better fits if we assume the capacitor has $4\%$ of the inductance.  The resonator impedance is around $30\textrm{-}35\;\Omega$. 

The devices were measured in an adiabatic demagnetization refrigerator with a base temperature of $50\;\rm{mK}$. The aluminum sample box holding the resonator chip was magnetically shielded with outer paramagnetic and inner niobium superconducting shields. The openings for the two microwave connectors were minimized to reduce incoming stray light. Measurements were performed with a vector network analyzer.  The measurement chain was comprised of a combination of attenuators (room temperature and cold) and a low pass filter on the input line to achieve the appropriate power level and eliminate radiation at the device input port (total attenuation -100 dB). On the output side of the sample box there was a microwave isolator and high- and low-pass filters before a high electron mobility transistor amplifier at the 4 K 2nd stage, with an additional room temperature amplifier. The power stored in the resonators was calculated in terms of both electic field and microwave photon numbers in the standard manner from the attenuation and measured resonance parameters \cite{Noroozian_AIP09}. Transmitted power measurements were taken as a function of frequency, power and temperature for all resonators. We measured over almost 4 decades of electric field (thus 7 decades in  photon numbers) and at low temperature, i.e. with a base temperature $T$ of $50\;\rm{mK}$, and a resonance frequency of $6 \;\rm{GHz}$, $T<\hbar\omega_r/2k_B$ and the TLSs are not thermally saturated.

\begin{figure}[tb]
\begin{center}
\includegraphics[width=8.5cm]{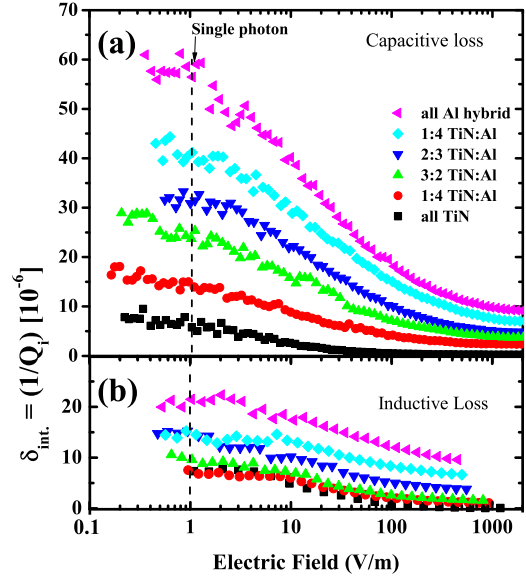}
\caption{(color online) Loss for (a) hybrid capacitor and (b) hybrid inductor resonators as a function of the electric field in the capacitor. The single photon levels are indicated.  When Al is added to the capacitor, there is a greater increase in loss than if Al is added to the inductor. }
\label{Fig_2}
\end{center}
\end{figure}

The electric field dependence of $\delta_\textrm{int.}$ at $50\;\rm{mK}$ is shown in Fig. \ref{Fig_2}. The intrinsic loss, $\delta_\textrm{int.}$, is constant for low eletric fields, before starting to drop off above the critical field. The loss is highest below the critical field, i.e. in the single photon regime, referred to here as $\delta({\textrm{SP}})$, shown at about 1 V/m, due to the presence of unsaturated TLSs. The highest loss was observed for the hybrid device with the all-Al capacitor, with  $\delta^{\Al}_{c}(SP)=55\pm5\times10^{-6}$, while the lowest loss was observed for the all-TiN devices, with $\delta^{\Ti\N}(SP)=5\pm1\times10^{-6}$. As expected, the loss curves of the hybrid Al-TiN devices fall between these two bounds. 

An alternative method of determining loss due to TLSs is done with a temperature sweep. In this method, $\delta^{\Ti\N}(TLS)$ is extracted at high power from the resonant frequency shift with temperature. This occurs because the off-resonant TLSs become thermally saturated. However, this technique can only be used when there are no other temperature dependent sources of loss, e.g. quasiparticles.  In particular, for Al, a superconductor with a relatively low $T_c$, the underlying condition, $T\ll T_c$ with resonator temperature $T$, does not hold. Hence, we can only use this technique for the all-$\Ti\N$ resonator. From this measurement, we obtained $\delta_{C}^{\Ti\N}\approx 5\pm1\times10^{-6}$, in agreement with the power-dependent loss shown in Fig. \ref{Fig_2}.

In quantum information applications, the very low electric field, i.e. single photon regime, loss is important.  However, this measured loss is composed of both power dependent loss terms, e.g. TLSs, as well as any power independent loss arsing from quasiparticles etc.  In order to separate these loss terms, we fit to the loss at the highest powers and subtract a constant loss such that the electric field dependence of loss matches equation (2), i.e. proportional to $1/E$ \cite{Pappas_ASC_11}.  We now have two loss terms, $\delta_{\textrm{TLS}}$ and $\delta_{\textrm{P.I.}}$, which we then can independently investigate the source of their loss.

If the TLS loss was solely in the inductor or capacitor, we would expect to only measure a change when $\Al$ is substituted in that element.  However, the circuit is not composed of completely lumped elements and there is finite capacitance in the meandered inductor amounting to  ~18$\%$ of the  total capacitance of the circuit.   When the loss is plotted vs $\Al$ capacitive fraction in Figure 3 (a), i.e. the 100$\%$ Al fraction hybrid C device has 82$\%$ $\Al$ capacitive fraction while the 100$\%$ Al fraction hybrid L device has 18$\%$ Al capacitive fraction, the extracted TLS losses for all devices show good agreement with a linear fit to the hybrid C data.  The hybrid L devices do exhibit slightly less loss than expected from the fit; but while the meandered inductor the same $2\;\rm{\mu m}$ spacing as the IDC, the exact electric field distribution and corresponding filling factor will be different for the meandered inductor versus the IDC and we should not expect perfect agreement. The fitted line from the hybrid Al devices closely intersects with the measured all-TiN devices indicating that any additional loss due to the TiN-Al interface is small, beneath the $\sim1\times10^{-6}$ uncertainty in our measurement.

The power independent loss term determined from the TLS fits increases with the amount of Al in the device, but is not proportional to either the amount of Al's inductance or capacitance.  However, if it is plotted versus the product of the (Al area)$\times$(Al inductance) as in Figure 3 (b) the loss is fit much better than to any single circuit parameter.  The small amount of inductance in the IDCs, $\sim4\%$  of the total,  is compensated by their 17$\times$ larger area than the meander.  The relationship between the power independent loss at the lowest temperatures indicates that its source is likely due to quasiparticles generated by stray IR light emanating from the 4K stage as reported recently in all -Al devices by other groups \cite{Corcoles_APL_2011,Barends_Stray_Light}.  Recent measurements with better absorptive materials in the box lead to a reduction in the power-independent loss term by up to $35\%$ for the all $\Al$ devices.

The loss in the hybrid devices can also be studied as a function of temperature.  In order to separate the temperature dependent loss from any potential TLS loss, the devices were measured at very high power where the TLS loss is saturated.  The measured temperature dependent losses are shown in Figure 4 (a).  As the temperature is increased in all devices, the loss rises exponentially. Devices with more Al have greater loss at all temperatures. This loss is enhanced when the Al is placed in the meandered inductor region.  In Figure 4 (b), the loss at 0.5 K in both hybrid L and C devices is plotted and agrees with a linear fit to the percentage of Al's inductance in the circuit. As the temperature dependent loss is in the inductance, it is likely due to thermally activated quasiparticles in the Al.

While the loss in the hybrid C devices at any particular temperature is proportional to the amount of Al, at higher temperatures the slope of the loss does not intercept the all TiN device at zero Al percentage, i.e. there exists additional loss at the interface between the TiN and the Al.  Figure 5 (a) shows the temperature dependent loss vs Al fraction for the hybrid C devices at temperatures betwen 0.2 and 0.9 K.  While the 5 devices with Al agree closely with linear fits (solid lines) to the Al fraction, the extrapolations of these fits (dashed lines) do not intersect with the all TiN resonator.  This implies that there is a relatively small additional loss term whenever Al is incorporated into the circuit, corresponding to the interface between the Al and the TiN.  This interfacial loss increases with temperature and is plotted in Figure 5 (b) along with an exponential fit the data.  The exponential fit with temperature suggests quasiparticles, and the activation energy scale is similar to the TiN superconducting gap energy.  While quasiparticles in the Al cannot transit to the TiN due to its larger energy gap, the opposite is not true.  Quasiparticles generated in the TiN can traverse the interface to the Al, be trapped there and generate additional loss perhaps by subsequently generating more quasiparticles in the Al. Presumably a similar interfacial effect would be visible in the hybrid L devices, but the comparatively small interface loss term would difficult to distinguish from the roughly $100\times$ larger loss in the Al inductors.

In summary, we have shown that the measured losses as a function of power and temperature can be varied by the replacement of Al for TiN in different parts of the resonant circuit.  When the lossier Al is placed in capacitive parts of the circuit, the increased TLS loss indicates that the TLSs couple to the capacitive parts of the circuit.  However while the Al in the inductor has much less TLS loss, it leads to additional power independent and temperature depenedent loss terms.  The power independent loss is proportional to the product of the Al area$\times$inductance and likely arises from quasiparticles generated by stray light from the 4 K stage.  The temperature dependent loss is related to the Al inductance and is likely from thermally activated quasiparticles.  While there is an incremental interface loss between the TiN and Al at higher temperatures, $\delta_{\textrm{interface}}\le 1\times10^{-6}$  at the base temperature.  This lack of interface loss does not preclude future hybrid devices from having low losses.  This work suggests that future devices can be designed with their individual circuit elements optimally tuned for their purpose e.g. low TLS loss, superconducting gap, kinetic inductance, detection efficiency etc. specifically tuned. \cite{MartinisPRL05,Weides_Trilayer,Weides_Transmon}

The authors gratefully acknowledge the assistance of F. Farhoodi and valuable discussions with J. Gao and J. Whittaker. This work was funded by the US government and supported by the NIST Quantum Information initiative. The views and conclusions contained in this document are those of the authors and should not be interpreted as representing the official policies, either expressly or implied, of the U.S. Government.

\newpage

\newpage
\begin{figure}[tb]
\begin{center}
\includegraphics[width=15cm]{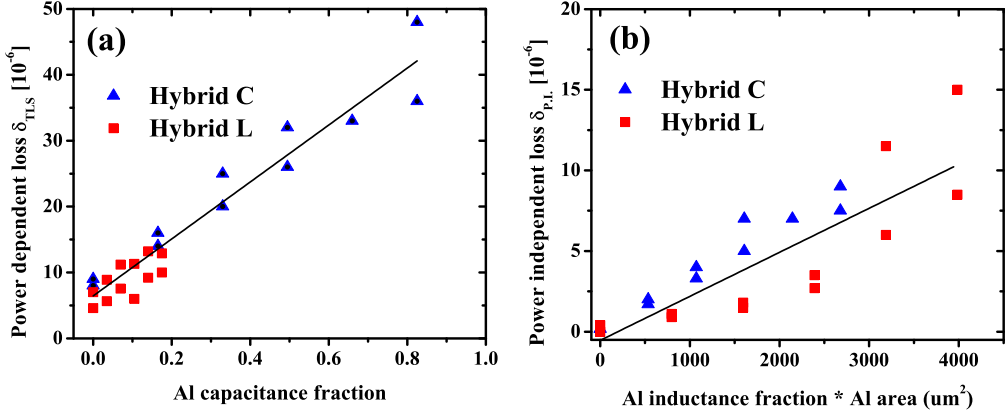}
\caption{(color online)  (a) The power dependent loss $\delta_{\textrm{TLS}}$ at 50 mK bath temperature versus Al capacitive fraction.  As the TiN is replaced by the Al in the hybrid L (square) or hybrid C (triangle) devices, additional power dependent loss is measured and agrees with a linear fit.  (b) The power independent loss, $\delta_{\textrm{P.I.}}$, versus the product of the Al inductance fraction$\times$area.  
}

\label{Fig_3}
\end{center}
\end{figure}

\newpage
\begin{figure}[tb]
\begin{center}
\includegraphics[width=15cm]{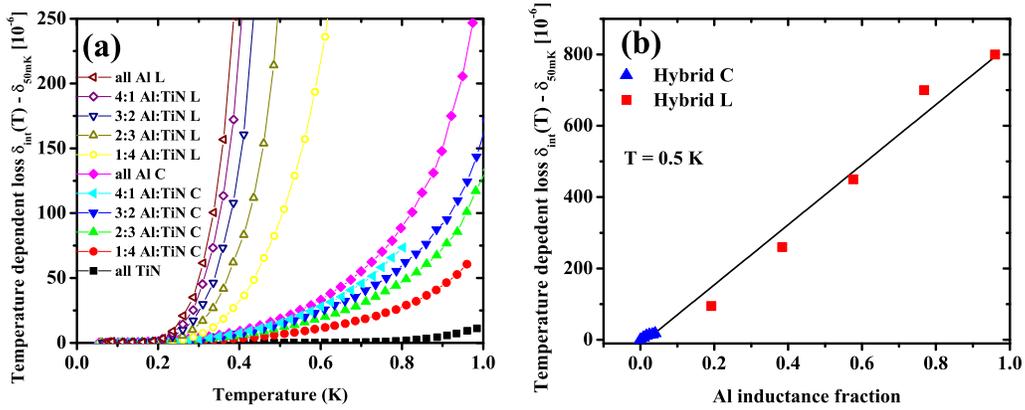}
\caption{ (color online) (a) Temperature dependent loss $\delta_{\textrm{int.}}(T)-\delta_{\textrm{50 mK}}$ of the hybrid C  (closed symbols) and hybrid L (open symbols) devices as a function of temperature.  The devices are measured at high power and the 50 mK bath temperature loss is subtracted to reduce the contribution of any other loss terms.  (b) The temperature dependent loss of all hybrid L and C resonators at 0.5 K versus Al inductance fraction agrees well to a linear fit.   }

\label{Fig_4}
\end{center}
\end{figure}

\newpage
\begin{figure}[tb]
\begin{center}
\includegraphics[width=15cm]{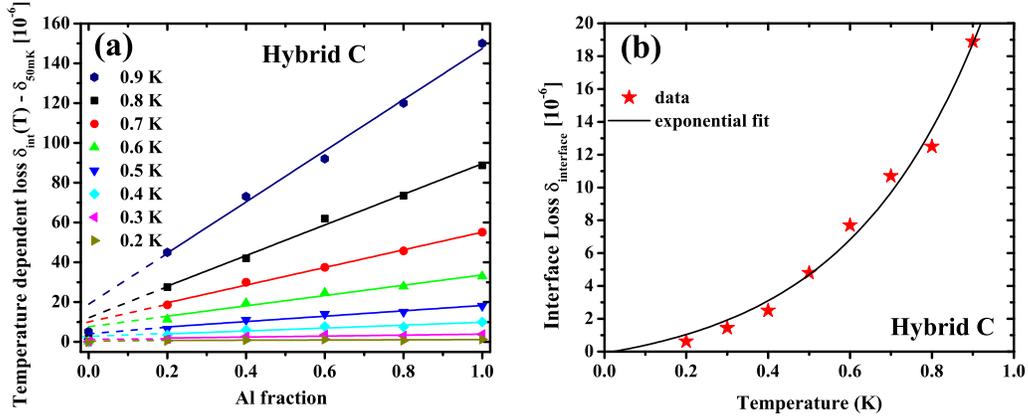}
\caption{(color online) (a) The temperature dependent loss $\delta_{\textrm{int.}}(T) -\delta_{\textrm{50 mK}}$ (symbols) in the hybrid C devices versus Al fraction at different temperatures between 0.2-0.9 K.  In the hybrid capacitor devices the fit (solid lines) to the temperature dependent loss of the 5 hybrid devices with Al is linear in Al fraction at all temperatures, but the extrapolations of these fits (dashed lines) do not intersect with the loss from the pure TiN device.  (b) The interface loss, $\delta_{\textrm{interface}}$ from the y-intercept from Fig. 5 a),  plotted vs temperature with an exponential fit (solid line). 
}
\label{Fig_5}
\end{center}
\end{figure}

\end{document}